\documentclass[%
 epj,
 sd,%
 amsmath,amssymb,
 reprint,%
]{revtex4-1}
\usepackage[english]{babel}
\usepackage{graphicx}
\usepackage{dcolumn}
\usepackage{bm}
\usepackage{color} 
\usepackage[caption=false]{subfig}

\begin{document}

\title{A Unified Description of Colloidal Thermophoresis}

\author{Jerome Burelbach}
\email{jb920@cam.ac.uk, ee247@cam.ac.uk}
\affiliation{Cavendish Laboratory, University of Cambridge, Cambridge CB3 0HE, United Kingdom}
\author{Daan Frenkel}
\affiliation{Department of Chemistry, University of Cambridge,  Cambridge, CB2 1EW, United Kingdom}
\author{Ignacio Pagonabarraga}
\affiliation{Departament de F\'isica de la Mat\`eria Condensada, Universitat de Barcelona, C. Mart\'i i Franqu\`es 1, 08028 Barcelona, Spain}
\affiliation{Institute of Complex Systems (UBICS), Universitat de Barcelona, Barcelona, Spain}
\affiliation{CECAM Centre Europ\'een de Calcul Atomique et Mol\'eculaire, \'Ecole Polytechnique F\'ed\'erale de Lausanne (EPFL), CH-1015 Lausanne, Switzerland}
\author{Erika Eiser}
\email{jb920@cam.ac.uk, ee247@cam.ac.uk}
\affiliation{Cavendish Laboratory, University of Cambridge, Cambridge CB3 0HE, United Kingdom.}

\begin{abstract}
We use the dynamic length and time scale separation in suspensions to formulate a general description of colloidal thermophoresis. Our approach allows an unambiguous definition of separate contributions to the colloidal flux and clarifies the physical mechanisms behind non-equilibrium motion of colloids. In particular, we derive an expression for the interfacial force density that drives single-particle thermophoresis in non-ideal fluids. The issuing relations for the transport coefficients explicitly show that interfacial thermophoresis has a hydrodynamic character that cannot be explained by a purely thermodynamic consideration. Our treatment generalises the results from other existing approaches, giving them a clear interpretation within the framework of non-equilibrium thermodynamics.
\end{abstract}

\maketitle

\section{Introduction}

The thermal motion of colloids in a temperature gradient is known
as thermophoresis. Since its discovery by Carl Ludwig and Charles
Soret in 1856 and in 1879 respectively \cite{Ludwig1856,Soret1879}, thermophoresis
has been studied experimentally in various systems, from charged particles in aqueous electrolyte solutions \cite{Putnam2005,Duhr2006b,Iacopini2006,Braibanti2008,Duhr2006,Dhont2007,Piazza2002,Piazza2003}
to long-chain polymers in polar or non-polar solvents \cite{Schimpf1987,Zhang1999,Duhr2004,Braun2002}.
Some of these studies have proven thermophoresis to be a promising
technique for the fractionation \cite{Jeon1997} or accumulation \cite{Duhr2006a}
of biomolecules. Thermophoresis is mainly governed by system-specific
interactions, which sometimes may be tuned such that different molecular
species migrate into opposite directions.

Although different models have already been proposed for colloidal
thermophoresis \cite{Alois2014,Demirel2001,Dhont2008,Dhont2004a,Parola2004}, a complete
theoretical description is still lacking. However, as the name suggests,
the consensus is that thermophoresis is a phoretic
phenomenon: the thermal motion of a colloid is mainly driven by local
hydrodynamic stresses in the surrounding liquid, confined in a region
close to the particle surface, often referred to as the interfacial
layer.

The flow of colloids in suspensions is quantified by the
net particle flux \cite{Piazza2008} 
\begin{equation}
\mathbf{J}=-D\nabla c-cD_{T}\nabla T,\label{eq:}
\end{equation}

where $D$ is the Fickian diffusion coefficient, $c$ is the colloidal concentration (number density),
$D_{T}$ is the thermal diffusion coefficient and T is the temperature.
The second term describes the particle flux induced by a temperature
gradient. From the relation $\mathbf{J}=c\mathbf{v}_{T}$, the thermophoretic
velocity can be identified as $\mathbf{v}_{T}=-D_{T}\nabla T$.

Most experimental techniques rely on observing the steady-state distribution
of colloids in a closed cell, which is reached when $\mathbf{J}=0$:
\begin{equation}
\nabla c=-cS_{T}\nabla T.\label{eq:-2}
\end{equation}

The ratio $S_{T}=\frac{D{}_{T}}{D}$ is called the Soret coefficient
and is widely used to quantify the strength of thermophoretic forces. From the definition of
$S_{T}$, it can be seen that colloids move to lower temperatures
if $S_{T}>0$ and to higher temperatures otherwise. Predicting the overall sign of $S{}_{T}$ is not trivial as thermophoresis
turns out to be an interplay of multiple contributions that may follow different trends \cite{Wurger2010}.

The difficulty in describing colloidal thermophoresis with a unique
theoretical model is twofold. First, colloidal masses and sizes are
much bigger than those of solvent molecules, but they are small enough
for the onset of Brownian motion. Secondly, thermophoresis is a non-equilibrium
phenomenon, meaning that a formulation based on local equilibrium thermodynamics only
applies under certain conditions \cite{deGroot1962}. Most theoretical models \cite{Parola2004,Ruckenstein1981,Dhont2007,Wurger2010} describe thermophoresis as driven by a gradient in surface tension or excess chemical potential, usually adopting either a purely hydrodynamic or thermodynamic viewpoint. In analogy to molecular thermodiffusion \cite{Alois2014,Wurger2014},
a thermodynamic approach relates the Soret coefficient to the excess enthalpy \cite{Wurger2013} or a gradient in thermodynamic potential \cite{Dhont2004a}, but it neglects dissipation via local fluid flows, thus restricting its validity to particles that are small compared to the interaction range. This dissipative character is correctly incorporated in a hydrodynamic approach \cite{Parola2004,Fayolle2008} that describes the fluid as a continuous medium subjected to stresses due to colloid-fluid interactions. However, hydrodynamic descriptions are usually formulated in a single-particle picture that ignores collective effects and Brownian motion.

So far, these approaches have mostly been discussed independently in literature due to a lack of common ground, although they are not mutually exclusive. This has lead to a general confusion and a disagreement about which thermophoretic contributions should be considered in a thermodynamic or hydrodynamic picture. Here, we show that the length and time scale separation in colloidal suspensions can be used to clarify this matter. This separation mainly occurs because the fluid particles are much smaller than the colloids and greatly exceed the colloids in number density. We derive system-specific relations between different transport coefficients that describe the coupling of thermodynamic forces to the colloidal flux. Our starting point is the theory of Non-Equilibrium Thermodynamics (NET), in which the temperature gradient is treated as a first order perturbation from equilibrium. NET has only received little attention in the discussion of colloidal thermophoresis, even though it provides a most general framework for thermal motion in multi-component systems.

\section{Non-Equilibrium Thermodynamics}

The theory of NET is based on the laws of thermodynamics, stating that the evolution of all components in a system is governed by its rate of entropy production. A key requirement for NET is that the system is at Local Thermodynamic Equilibrium
(LTE), meaning that it can be partitioned into small volume
elements, each of which may be assumed in thermodynamic equilibrium.
This condition is usually satisfied for moderate temperature gradients
in the absence of large-scale advection \cite{Duhr2006,deGroot1962}. An important thermodynamic relation that remains valid for a volume element at LTE is the Gibbs-Duhem equation \cite{deGroot1962} 
\begin{equation}
dP=sdT+\sum_{k}n_{k}d\mu_{k},\label{eq:-17}
\end{equation}

where $s$ is the entropy density and $P$ is the total pressure of the volume element. $n_k$ is the number concentration of component $k$ and $\mu_k$ is the corresponding chemical potential. In the presence of thermodynamic gradients, the Gibbs-Duhem equation can be interpreted as a balance equation for the forces acting on a local volume element.

Let us now consider a continuous thermodynamic system at LTE, in the absence of chemical
reactions. From the resulting balance equations for heat, mass and internal energy, it can be shown that the rate of entropy production
$\sigma_{s}$ inside a volume element takes the following form \cite{deGroot1962}:

\begin{equation}
\sigma_{s}=\mathbf{J}_{q}\nabla\frac{1}{T}+\sum_{k}\mathbf{J}_{k}\left\{ -\nabla\frac{\mu_{k}}{T}+\frac{1}{T}\mathbf{F}_{k}\right\} -\frac{1}{T}\Gamma:\nabla\mathbf{u},\label{eq:-9}
\end{equation}

where $\Gamma$ is the viscous stress tensor and $\mathbf{u}$
is the centre of mass velocity of the volume element. $\mathbf{J}_{k}=n_k\left(\mathbf{v}_k-\mathbf{u}\right)$ is the net particle
flux of component $k$ relative to $\mathbf{u}$, satisfying $\sum_{k}m_{k}\mathbf{J}_{k}=0$, where $m_k$ is the corresponding particle mass.
The total heat flux $\mathbf{J}_{q}$ accounts for both heat conduction and
heat diffusion and the body force $\mathbf{F}_{k}$ includes external
forces as well as internal forces whose range exceeds the typical LTE scale (\textit{e.g.} thermoelectric forces). A more convenient form of eq. (\ref{eq:-9}) can be obtained by rewriting $\nabla\frac{\mu_{k}}{T}$ as

\begin{equation}
\nabla\frac{\mu_{k}}{T}=\bar{H}_k\nabla\frac{1}{T} + \frac{1}{T}\nabla_T\mu_k,\label{eq:-45}
\end{equation}

where

\begin{equation}
\bar{H}_k=-T^2\frac{\partial}{\partial T}\left(\frac{\mu_{k}}{T} \right)_{P,n_j}\label{eq:-46}
\end{equation}

is the partial molar enthalpy of component $k$. With eq. (\ref{eq:-45}), the rate of entropy production can now be expressed as

\begin{equation}
\sigma_{s}=\mathbf{J}'_{q}\nabla\frac{1}{T}+\frac{1}{T}\sum_{k}\mathbf{J}_{k}\left\{ -\nabla_T\mu_{k}+\mathbf{F}_{k}\right\} -\frac{1}{T}\Gamma:\nabla\mathbf{u},\label{eq:-48}
\end{equation}

where the 'modified' heat flux $\mathbf{J}'_{q}$ is related to $\mathbf{J}_{q}$ via

\begin{equation}
\mathbf{J}'_{q}=\mathbf{J}_{q}-\sum_k \bar{H}_k\mathbf{J}_{k}.\label{eq:-69}
\end{equation}

Eq. (\ref{eq:-48}) shows that entropy can be produced by two vectorial fluxes $\mathbf{J}'_{q}$ and $\mathbf{J}_{k}$; and one tensorial flux related to the fluid flow gradient $\nabla\mathbf{u}$. Onsager's theory of NET postulates linear constitutive relations between the vectorial fluxes and thermodynamic forces, of the form

\begin{eqnarray}
\mathbf{J}_{i} & = & L_{iq}\nabla\frac{1}{T}+\frac{1}{T}\sum_{k}L_{ik}\left\{ -\nabla_T\mu_{k}+\mathbf{F}_{k}\right\}, \label{eq:-10}\\
\mathbf{J}'_{q} & = & L_{qq}\nabla\frac{1}{T}+\frac{1}{T}\sum_{k}L_{qk}\left\{ -\nabla_T\mu_{k}+\mathbf{F}_{k}\right\}, \label{eq:-11}
\end{eqnarray}

where the scalar coefficients $L$ are known as the Onsager transport coefficients. The flux induced by an external force $\mathbf{F}_{i}$ is more commonly written as

\begin{equation}
\mathbf{J}_{i}=\frac{n_{i}}{\xi_{i}}\mathbf{F}_{i},
\end{equation}

where $\xi_{i}$ is the friction coefficient of a particle of component $i$. As a result, $\xi_{i}$ and $L_{ii}$ are related by

\begin{equation}
L_{ii} = \frac{n_{i}T}{\xi_{i}}.
\end{equation}

An important feature of Onsager's theory, also known as the reciprocal relations, is that the cross-coefficients are symmetric, so that \cite{Onsager1931,Onsager1931a}

\begin{equation}
L_{ik}=L_{ki}\hspace{0.3cm} \text{and} \hspace{0.3cm}L_{iq}=L_{qi}.
\end{equation}

Although the Curie symmetry principle forbids coupling between tensorial forces and vectorial fluxes in a homogeneous isotropic medium, a local hydrodynamic coupling between shear flows and vectorial forces can occur inside the interfacial layer around a colloid. Furthermore, it should be noted that the Onsager flux (\ref{eq:-10}) carries a large number of
variables in an $N$-component system, with $(1+N)N/2$ independent transport coefficients and $N-1$
independent thermodynamic forces. This suggests that an introduction of specific assumptions is required to achieve a hydrodynamic description of thermophoresis in terms of a reduced number of independent variables.

\section{Dynamic Length and Time Scale Separation in Colloidal Suspensions}

Onsager's theory provides general expressions for the particle and heat fluxes, but it makes no attempt to determine the relevant transport coefficients $L$ in specific thermodynamic systems. Here, we construct a framework that allows the formulation of system-specific relation between these coefficients for thermophoresis in colloidal suspensions. The system of interest is a closed suspension at LTE, subjected to a constant and uniform temperature gradient by keeping opposite sides of the system in contact with thermostats at different temperatures. It is assumed that the system is not subjected to any external forces, so that the total pressure $P$ of the system is uniform everywhere. The colloids are dispersed in a fluid that mainly consists of solvent molecules, but that can additionally contain small solutes of negligible size (\textit{e.g.} ions). In the following, the index $i=0$ is reserved for the solvent. The colloidal concentration and flux are denoted by $c$ and $\mathbf{J}$ respectively, and the index $i=1$ is used to refer to other quantities of the colloidal component.  

Our framework is based on the dynamic length and time scale separation between the colloid and fluid \cite{Piazza2003,Brady2011} and we therefore introduce the following assumptions:

\begin{enumerate}
	\item The colloids are much larger/heavier than fluid particles
	\item The component densities satisfy $c\ll n_{k\neq 0,1}\ll n_0$
	\item The solvent is incompressible 
	\item Fluid flow has a Reynolds number much smaller than one 
	\item Fluid mass diffusion dominates over fluid advection and colloidal motion (the fluid Peclet number is much smaller than one)
\end{enumerate}

This set of assumptions forms the basis for the hydrodynamic approach to thermophoresis. In particular, assumptions 1 and 2 allow the use of the continuum approximation. The fluid may thus be treated as a continuous medium and the incompressibility of the solvent allows an 'instantaneous' equilibration of the pressure $P$, such that $\nabla P=0$. Further, the presence of a large bulk reservoir of pure fluid allows the introduction of an effective bulk fluid pressure $P_s^b$, which can be defined via eq. (\ref{eq:-17}) as the pressure resulting from thermodynamic forces inside a volume element of pure fluid:

\begin{equation}
dP_s^b=s_s^bdT+\sum_{k\neq 1}n_{k}^bd\mu_{k},\label{eq:-20}
\end{equation}

where $s_s^b$ is the entropy density of the bulk fluid and $n_{k}^b$ is the corresponding bulk concentration of component $k$. 

For colloids, a departure from the ideal state occurs due to specific interactions with the surrounding components. The colloidal chemical potential can then more generally be written as $ \mu_{1}=\mu_{id} + \mu_{exc}$, where $\mu_{id}$ is the ideal chemical potential. The excess chemical potential $\mu_{exc}$ accounts for a specific interaction between colloid and fluid, denoted by $\mu_{cs}$; and
for a collective contribution $\mu_{cc}$ due to hard-core interactions or specific pair-interactions between colloids. According to assumptions 1 and 5, the fluid responds to these interactions with a rapid relaxation to a local equilibrium distribution around the colloids that remains unperturbed by colloidal motion or advection. At uniform temperature, this allows the formulation of a 'reduced' description \cite{Dijkstra1999,Felderhof2003}, in which the colloid-fluid interaction $\mu_{cs}$ is treated as a local interfacial layer around the colloid, separated out from the bulk. Inside the interfacial layer, the local thermodynamic properties of the fluid differ from those of the bulk fluid, which in turn barely feels the presence of the colloids. As the introduction of a colloid necessarily leads to the build-up of an interfacial layer, $\mu_{cs}$ is equal to the surface energy of the created interface: 

\begin{equation}
\mu_{cs}=A_c\left( \frac{\partial G}{\partial A}\right)_{P,T,N_{k\neq 1}} =A_c\gamma_{cs},\label{eq:-31}
\end{equation}

where $\gamma_{cs}$ is the interfacial tension.The surface area $A_c$ is assumed constant, meaning that the increase in surface area $\partial A$ exclusively occurs by adding colloids to the suspension. The change in surface energy can further be related to interfacial excess properties of the fluid via the Gibbs adsorption equation

\begin{equation}
-d\mu_{cs}= S_{\phi}dT+\sum_{k\neq1}N_{k}^{\phi}d_T\mu_{k},\label{eq:-56}
\end{equation}

where $N_{k}^{\phi}$ is the excess number of fluid particles of component $k$ and $S_{\phi}$ is the interfacial excess entropy. As equal and opposite forces are exerted on the colloid and its interfacial layer, using eq. (\ref{eq:-56}) at uniform temperature further yields the relation

\begin{equation}
-\nabla_T\mu_{cs}+\mathbf{F}_{1}=-\sum_{k\neq 1}N_{k}^{\phi}\left\{-\nabla_T\mu_{k}+\mathbf{F}_{k}\right\}.\label{eq:-55} 
\end{equation}

A collective contribution $\mu_{cc}$ arises from the interaction between overlapping layers. From this description, it follows that the colloidal chemical potential can be expressed as a sum of two separate terms: 

\begin{equation}
\mu_{1}=\mu_{cs} + \mu_{c},\label{eq:-1}
\end{equation}

where $\mu_{c}=\mu_{id}+\mu_{cc}$ is the 'bulk' chemical potential of the colloidal component. To make progress in the description of thermophoresis, we assume that this superposition principle can be extended to colloidal motion in a temperature gradient,  so that the total flux can be written as

\begin{equation}
\mathbf{J}=\mathbf{J}_{cs}+\mathbf{J}_{c}. \label{eq:-7}
\end{equation}

This is achieved by formulating a source term analogous to eq. (\ref{eq:-9}) separately for the bulk entropy of the suspension and the excess entropy of the fluid at the colloidal surface \cite{Kjelstrup2008}. However, the separation of the flux in eq. (\ref{eq:-7}) also relies on the fact that the hydrodynamic flows induced by each term can be treated as decoupled from each other. This assumption is indeed valid for low Reynolds number fluids, where the linear Stokes equation allows the use of the superposition principle of fluid flows. A similar separation must also hold for the heat fluxes, implying that heat transport inside the interfacial layer is predominantly due to interfacial flows, which in turn do not significantly contribute to the transport of heat in the bulk of the suspension. Following these arguments, $\mathbf{J}_{c}$ and $\mathbf{J}_{cs}$ can now be written as two decoupled Onsager fluxes 

\begin{equation}
\mathbf{J}_{c}= L_{1q}^{c}\nabla\frac{1}{T}-\frac{L_{11}}{T}\nabla_T\mu_{c}+\frac{1}{T}\sum_{k\neq1}L_{1k}^c\left\{ -\nabla_T\mu_{k}+\mathbf{F}_{k}\right\}\label{eq:-23}
\end{equation}

and

\begin{eqnarray}
\mathbf{J}_{cs} & = &  L_{1q}^{cs}\nabla\frac{1}{T}+\frac{L_{11}}{T}\left( -\nabla_T\mu_{cs}+\mathbf{F}_{1}\right)\nonumber\\
& & +\frac{1}{T}\sum_{k\neq1}L_{1k}^{cs}\left\{ -\nabla_T\mu_{k}+\mathbf{F}_{k}\right\}.\label{eq:-18}
\end{eqnarray}

For the hydrodynamic considerations that are about to follow, it is useful to eliminate the term $-\nabla_T\mu_{cs}+\mathbf{F}_1$ with  eq. (\ref{eq:-55}), allowing us to express eq. (\ref{eq:-18}) in the alternative form

\begin{equation}
\mathbf{J}_{cs}=\frac{L_{11}}{T}\left( -Q_{cs}^*\frac{\nabla T}{T}+\sum_{k\neq1}N_{k}^*\left\{ -\nabla_T\mu_{k}+\mathbf{F}_{k}\right\}\right), \label{eq:-57}
\end{equation}

where the interfacial excess quantities $Q_{cs}^*$ and $N_k^*$ are given by

\begin{eqnarray}
Q_{cs}^* & = & L_{1q}^{cs}/L_{11},\label{eq:-27}\\
N_{k}^* & = & L_{1k}^{cs}/L_{11}-N_{k}^{\phi}.\label{eq:-74}
\end{eqnarray}

A carefully chosen set of assumptions that specifically applies to colloidal suspensions has thus led us to a framework in which the separate evaluation of $\mathbf{J}_{cs}$ and $\mathbf{J}_{c}$ is reasonable. As a result, the interfacial contribution $\mathbf{J}_{cs}$ may now be determined in a hydrodynamic single-particle picture, which is the subject of the next section.

\section{The Interfacial Contribution: Hydrodynamic Approach}

\begin{figure}
\centering{}\includegraphics[width=6.4cm,height=6cm]{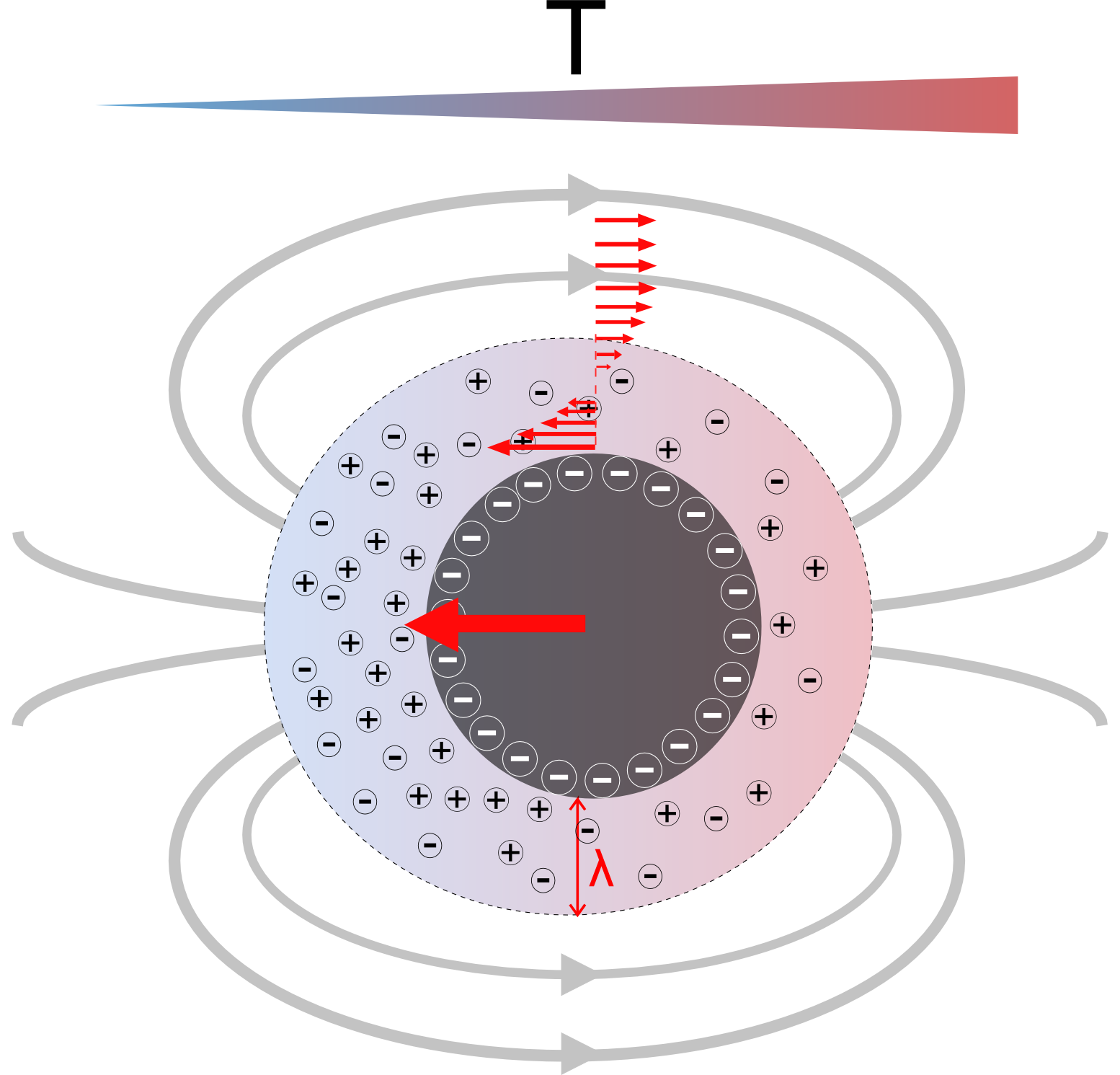}\caption{Schematic depiction of hydrodynamic stresses caused by a temperature
gradient inside the electric double layer around a charged colloid.
The gradient in excess pressure induces a thermo-osmotic
flow close to the colloidal surface (grey lines). In response, the
colloid moves in the opposite direction (big red arrow).\label{fig:} }
\end{figure}

The hydrodynamic picture discusses how thermodynamic bulk gradients induce interfacial
stresses in the fluid close to the surface of a single colloid, by treating the colloid as
a macroscopic object and the surrounding fluid as a continuous medium.
It is well known that a thermodynamic gradient across an interfacial layer gives rise to an interfacial fluid flow in one direction and a corresponding phoretic drift of the colloid in the opposite direction \cite{Anderson1989} (Fig. \ref{fig:}). In a homogeneous system at uniform temperature, a radially symmetric distribution of fluid around the colloid is maintained
by a local balance between a body force density $\mathbf{f}$ and a gradient in fluid pressure
$P_s$, such that $\mathbf{f}-\nabla P_s=0$. A thermodynamic bulk gradient (in temperature or chemical potential) then breaks this balance and sets the colloid and fluid into motion. A steady-state drift velocity $\mathbf{v}$ is reached when the total force on the colloid is zero and the resulting colloidal flux can then be written as

\begin{equation}
\mathbf{J}_{cs}=c\mathbf{v}=\frac{c}{\xi}\mathbf{F}_{cs},\label{eq:-49} 
\end{equation}

where, in view of eq. (\ref{eq:-57}), the interfacial driving force $\mathbf{F}_{cs}$ is given by

\begin{equation}
\mathbf{F}_{cs}=-Q_{cs}^*\frac{\nabla T}{T}+\sum_{k\neq1}N_k^*\left\{ -\nabla_T\mu_{k}+\mathbf{F}_{k}\right\}.\label{eq:-73}
\end{equation} 

Although the 'interfacial heat of transport' $Q_{cs}^*$ has commonly been identified as the driving force behind interfacial thermophoresis, the contribution related to $N_{k}^*$ has often been overlooked. This is rather surprising, as it is the latter contribution that can give rise to the well-known effect of diffusiophoresis at uniform temperature. Based on Onsager's reciprocal relations, $Q_{cs}^*$ and $N_k^*$ can be determined from the heat and particle fluxes that arise inside the interfacial layer when the colloid moves through a homogeneous fluid at uniform temperature. The corresponding interfacial excess densities of the fluid must however be defined carefully before these fluxes can be computed. For this purpose, we first consider the momentum balance equation of the fluid, which is governed by the Navier-Stokes equation 

\begin{equation}
\vec{\mathcal F}+\eta\nabla^{2}\mathbf{u}_s=0,\label{eq:-44}
\end{equation}

where $\vec{\mathcal F}=\mathbf{f}-\nabla P_s$ is the net force density acting on a fluid element, $\mathbf{u}_s$ is the local centre of mass velocity of the fluid and $\eta$ is the fluid viscosity. The inertia term has been neglected in eq. (\ref{eq:-44}) due to the assumption of small Reynolds number. Our aim is to derive a general expression for the excess force density $\vec{\mathcal F}_{\phi}$ that drives interfacial fluid flow. In recent literature
\cite{Wurger2010,Dhont2008,Parola2004}, different expressions have only been given in the limit where the interfacial excess of fluid is described by Poisson-Boltzmann theory, suggesting that a general expression of $\vec{\mathcal F}_{\phi}$ for non-ideal fluids is still lacking.

We start by considering a colloid whose surface is in contact with a fluid made of solvent molecules and small solutes. The solvent is pictured as an incompressible, polarisable medium. Due to the linearity of eq. (\ref{eq:-44}), the interfacial force density $\vec{\mathcal F}_{\phi}$ can be treated as decoupled from the subsequent stresses induced by collective colloidal motion. In the following, we denote a fluid property $x$ with an index $b$ to refer to its value in the bulk and by $x(\mathbf{r})$ to refer to its local value at a position $\mathbf{r}$ from the colloidal centre. Excess densities will be denoted with an index $\phi$, to show that they rely on the presence of a specific interaction between colloid and fluid. A fluid component $k$ can be subjected
to a local conservative body force $-\nabla_T\phi_{k}(\mathbf{r})$ , deriving from
a potential $\phi_{k}$ at the colloidal surface that tends to zero
in the bulk; and a body force $\mathbf{F}_{k}$ induced by the temperature
gradient in the bulk. The local body force density on a fluid element
is thus given by
\begin{equation}
\mathbf{f}=-\sum_{k\neq1}n_{k}(\mathbf{r})\left(\nabla_T\phi_{k}(\mathbf{r})-\mathbf{F}_{k}\right)+\mathbf{p}\nabla\mathbf{E}(\mathbf{r}),\label{eq:-37}
\end{equation}

where the last term accounts for the electric force due to the solvent
polarization $\mathbf{p}$ in the non-uniform electric field $\mathbf{E}$ of the colloid.
Further, the Gibbs-Duhem equation for a polarisable medium can be used to relate the gradient in fluid pressure $P_s$ to thermodynamic gradients at the colloidal surface \cite{deGroot1962}: 

\begin{equation}
\nabla P_s=s_s(\mathbf{r})\nabla T+\sum_{k\neq1}n_{k}(\mathbf{r})\nabla\mu_{k}(\mathbf{r})+\mathbf{p}\nabla\mathbf{E}(\mathbf{r}).\label{eq:-35}
\end{equation}

In order to express eq. (\ref{eq:-35}) in terms of the same thermodynamic forces as eq. (\ref{eq:-57}), we split $\nabla\mu_k(\mathbf{r})$ up into

\begin{equation}
\nabla\mu_{k}(\mathbf{r})=-\bar{S}_k\nabla T + \nabla_T\mu_k(\mathbf{r}),
\end{equation}

where $\bar{S}_k$ is the partial molar entropy of component $k$. Substitution into eq. (\ref{eq:-65}) then yields 

\begin{equation}
\nabla P_s=s'_s(\mathbf{r})\nabla T+\sum_{k\neq1}n_{k}(\mathbf{r})\nabla_T\mu_{k}(\mathbf{r})+\mathbf{p}\nabla\mathbf{E}(\mathbf{r}),\label{eq:-65}
\end{equation}

where $s'_s(\mathbf{r})$, the 'modified' contribution to $s(\mathbf{r})$, is given by

\begin{equation}
s'_s(\mathbf{r})=s_s(\mathbf{r})-\sum_{k\neq1}n_{k}(\mathbf{r})\bar{S}_k.\label{eq:-71}
\end{equation}

The 'modified' contributions related to other extensive thermodynamic quantities can be defined analogously and are henceforth denoted with a prime. It is crucial to note the delicate difference between the entropy densities $s'_s$ and $s_s$. The change from $s_s$ to $s'_s$ is analoguous to the transition from $\mathbf{J}_q$ to $\mathbf{J}'_q$, which naturally arises when the basis of thermodynamic forces is changed from ($\nabla\frac{1}{T}$,$\nabla\mu_k$) to the linearly independent set ($\nabla\frac{1}{T}$,$\nabla_T\mu_k$). A discussion of entropy and heat flux is therefore only meaningful if these quantities are clearly specified within the chosen basis.

With eqs. (\ref{eq:-65}) and (\ref{eq:-37}), the local force density $\vec{\mathcal F}=\mathbf{f}-\nabla P_s$ on a fluid element equals 
\begin{equation}
\vec{\mathcal F}=-s'_s(\mathbf{r})\nabla T-\sum_{k\neq1}n_{k}(\mathbf{r})\left\{ \nabla_T\left( \mu_{k}(\mathbf{r})+\phi_{k}(\mathbf{r})\right) -\mathbf{F}_{k}\right\}. \label{eq:-5}
\end{equation}

In a homogeneous system at uniform temperature, the equilibrium structure of the interfacial layer around a colloid is determined by the condition of zero force density 

\begin{equation}
\vec{\mathcal F}=-\sum_{k\neq1}n_{k}(r)\nabla_T\left(\mu_{k}(r)+\phi_{k}(r)\right)=0,
\end{equation}

where all quantities only depend on the radial distance $r$ from the colloidal centre due to the radial symmetry. This condition is satisfied if $\nabla_T\left(\mu_{k}(r)+\phi_{k}(r)\right)=0$.
Integration from the colloidal surface into the bulk of the suspension then directly yields

\begin{equation}
\mu_{k}(r)+\phi_{k}(r)=\mu_{k}^{b},\label{eq:-82}
\end{equation}

where $\mu_{k}^{b}$ is the chemical potential of component $k$ in the bulk. In a non-equilibrium system, $\mu_{k}^{b}$ can more generally be understood as the value of the chemical potential far away from the colloidal surface, along the isotherm of the considered fluid element. 

The condition of LTE implies that the chemical equilibrium given by eq. (\ref{eq:-82}) remains valid in a temperature gradient when the temperature $T$ is approximately constant over the layer. Within the scope of NET, the net force on the colloid is thus evaluated to first in the gradients by assuming that the interfacial layer remains radially symmetric. This crucial assumption further allows us to redefine the fluid chemical potential $\mu_{k}$ by including the potential $\phi_k$ as an internal interaction in the fluid equation of state:

\begin{equation}
\mu_{k}\equiv\mu_{k}(r)+\phi_{k}(r)=\mu_{k}^{b}, \hspace{0.5cm}\bar{S}_k=-\left( \frac{\partial\mu_{k}}{\partial T}\right) _{P,n_j}
\end{equation}

The index '$b$' for $\mu_k$ can hence simply be omitted and it directly follows from the standard relations $Ts=h-\sum n_k\mu_k$ and $T\bar{S}_k=\bar{H}_k-\mu_k$ that

\begin{equation}
Ts'_{s}(r)=h'_{s}(r),\label{eq:-16}
\end{equation}

where $h'_{s}(r)$ is the corresponding 'modified' contribution to the fluid enthalpy density. Further, eq. (\ref{eq:-5}) can now be written as 

\begin{equation}
\vec{\mathcal F}=-h'_s(r)\frac{\nabla T}{T}+\sum_{k\neq1}n_{k}(r)\left\{-\nabla_T\mu_{k} +\mathbf{F}_{k}\right\},\label{eq:-12}
\end{equation}

where $h'_s(r)$ and $n_{k}(r)$ only depend on the radial distance $r$ from the colloidal centre. As interfacial thermophoresis is concerned with the part of $\vec{\mathcal F}$ resulting from the specific interaction between colloid and fluid, we have to subtract from eq. (\ref{eq:-12}) the value of $\vec{\mathcal F}$ in the absence of the interfacial layer, giving

\begin{equation}
\vec{\mathcal F}_{\phi}=-q_\phi(r)\frac{\nabla T}{T}+\sum_{k\neq1}n_{k}^{\phi}(r)\left\{ -\nabla_T\mu_{k}+\mathbf{F}_{k}\right\}\label{eq:-14}
\end{equation} 

with 

\begin{eqnarray}
q_\phi(r) & = & h_{\phi}(r)=h'_s(r)-h_{s}'^{b},\label{eq:-22}\\
n_{k}^{\phi}(r) & = & n_{k}(r)-n_{k}^{b},\label{eq:-60}
\end{eqnarray}

where $q_\phi(r)$ is the interfacial heat density and $n_{k}^{\phi}(r)$ is the interfacial excess (number) density of fluid component $k$. 

With eq. (\ref{eq:-14}), we have thus derived a most general expression for the excess force density resulting from the specific interaction between colloid and fluid. This result specifically relies on the assumption of LTE inside the interfacial layer and shows that thermodynamic forces couple to the interacial excess densities of the fluid, which are now unambiguously defined by eqs. (\ref{eq:-22}) and (\ref{eq:-60}). It should however be noted that eq. (\ref{eq:-14}) ignores heat conduction through the colloid, which must be taken into account if its thermal conductivity $\kappa_{c}$ differs from the conductivity $\kappa_{s}$ of the fluid. For convenience, let us denote the interfacial excess densities ($q_{\phi}$ or $n_{k}^{\phi}$) by $x_{\phi}$ and the corresponding interfacial excess quantities ($Q_{cs}^*$ or $N_k^*$) by $X^*$. Based on Onsager's reciprocal relations, the general form of $X^*$ can be obtained by noticing that the integrated flux $X^*\mathbf v$ resulting from the 'interfacial polarization' of a colloid moving with a velocity $\mathbf v$ through a homogeneous fluid at uniform temperature is given by \cite{Agar1989}

\begin{equation}
X^*\mathbf v=\int_{R}^{\infty}x_{\phi}(r)\mathbf u_s\left(\mathbf r \right)dV,\label{eq:-13}
\end{equation}

where $R$ is the radius of the colloid and $\mathbf u_s\left(\mathbf r \right)$ is the induced fluid flow inside the rest frame of the colloid.
As the interfacial excess density $x_{\phi}(r)$ only depends on radial distance, the angular integration in eq. (\ref{eq:-13}) can be carried out over the fluid flow, yielding (see Appendix A)

\begin{equation}
X^*=-\int_{R}^{\infty}4\pi r^2\left(1-b\frac{R}{r}\right)x_{\phi}(r)dr,\label{eq:-54}
\end{equation}

where the dimensionless constant $b$ takes the value $b=1$ for stick and $b=2/3$ for slip boundary conditions at the colloidal surface.

Now, let us further introduce a characteristic length scale $\lambda$ that defines the 'thickness' of the interfacial layer. Of particular interest are the limiting cases of
'large layers' ($R\ll\lambda$) and 'thin layers' ($R\gg\lambda$), which are respectively known as the H\"uckel limit \cite{Morthomas2008} and the boundary layer approximation \cite{Wurger2010}. In the H\"uckel limit, the particle size is negligible ($R/r\rightarrow 0$) and eq. (\ref{eq:-54}) reduces to a volume integral over the layer. Further, heat conduction through the colloid can be ignored, so that $q_{\phi}=h_{\phi}$. We thus obtain

\begin{equation}
\mathbf{F}_{cs}= H_{\phi}\frac{\nabla T}{T}-\sum_{k\neq1}N_{k}^{\phi}\left\{ -\nabla_T\mu_{k}+\mathbf{F}_{k}\right\},\label{eq:-78}
\end{equation}

where $N_k^{\phi}=\int n_k^{\phi}dV$ is the interfacial excess of fluid particles and $H_{\phi}=\int h_{\phi}dV$ is the interfacial excess enthalpy. The flux $\mathbf{J}_{cs}$ is hence independent of the boundary condition at the colloidal surface and the corresponding Onsager coefficients reduce to

\begin{eqnarray}
L_{1q}^{cs} & = & -H_{\phi}L_{11},\\
L_{1k}^{cs} & = & 0.
\end{eqnarray}

Further, eqs. (\ref{eq:-56}) and (\ref{eq:-55}) can be used to rewrite eq. (\ref{eq:-78}) in the alternative form

\begin{equation}
\mathbf{F}_{cs}= -\nabla\mu_{cs}+\mathbf{F}_1. \label{eq:-24}
\end{equation}

This result shows that the H\"uckel limit corresponds to an effective 'thermodynamic' treatment of colloidal motion, driven by a gradient in surface energy $-\nabla\mu_{cs}$. As the H\"uckel limit is restricted to particles that are small compared to the layer thickness, it is however not expected to hold for colloidal thermophoresis. Colloids usually have diameters that largely exceed the interaction range and should therefore be considered in the boundary layer approximation ($R\gg\lambda$), where the heat flux through the colloid modifies its thermal polarization. In this limit, the interfacial heat density is therefore no longer equal to the interfacial enthalpy density $h_\phi$ but can be related to $h_\phi$ via $q_\phi=Ch_\phi$ where the constant $C$ is set by the ratio between $\kappa_{c}$ and $\kappa_{s}$ \cite{Wurger2010}. Alternatively, one can directly derive a similar relation between the integrated heat and enthalpy flux, as shown in Appendix B. By expanding eq. (\ref{eq:-54}) to first order in $z/R\ll 1$ where $z=r-R$ is the distance from the colloidal surface, we find:

\begin{eqnarray}
X^* & = & -4\pi R^2(1-b)\int_{0}^{\infty}x_{\phi}(z)dz\\
& & \hspace{2cm}-4\pi R(2-b)\int_{0}^{\infty}zx_{\phi}(z)dz, \label{eq:-3}\nonumber\\
& = & 
\begin{cases}
-4\pi R\int_{0}^{\infty}zx_{\phi}(z)dz& \text{for stick}\\
\\
-\frac{1}{3}X_{\phi}-\frac{16}{3}\pi R\int_{0}^{\infty}zx_{\phi}(z)dz & \text{for slip,}
\end{cases}
\end{eqnarray}

where $X_{\phi}=\int x_{\phi}dV=4\pi R^2\int x_{\phi}dz$. 

Interestingly, the expression for a stick boundary in eq. (\ref{eq:-3}) coincides with the expression first derived by Derjaguin, who based his derivation on Onsager reciprocity by considering isothermal fluid flow through a porous medium \cite{Derjaguin1987,Anderson1989}. An important feature of the boundary layer approximation is that, although thermophoretic motion is still \textit{induced} by a gradient in surface energy $\mu_{cs}$, the force $\mathbf{F}_{cs}$ that \textit{drives} thermophoresis can no longer just be written as $-\nabla\mu_{cs}$. In general, we note that this hydrodynamic nature of thermophoresis is characterised by a non-zero coupling coefficient $L_{1k}$ and a value of $-Q_{cs}^*$ that differs from the interfacial excess enthalpy $H_{\phi}$. It can further be seen from eq. ( \ref{eq:-54}) that the thermodynamic limit ($R/r\rightarrow 0$) constitutes an upper bound for $\mathbf{F}_{cs}$. As a result, the presence of a solid surface generally leads to dissipative effects that tend to inhibit thermophoretic motion.

\section{The Bulk Contribution: Collective Effects}

We now turn to the remaining bulk contribution $\mathbf J_c$ that represents the effect of Brownian motion and collective effects. Collective thermophoresis is usually described using a microscopic approach that relies on a clear separation between inter-colloidal and interfacial interactions. To justify the validity of such an approach, let us first consider the Gibbs-Duhem equation for a volume element at LTE:

\begin{equation}
c\nabla_T\mu_{1}+\sum_{k\neq 1}n_{k}\nabla_T\mu_{k}=0.\label{eq:-43}
\end{equation}

In order to obtain a balance equation for the bulk of the suspension, we need to make eq. (\ref{eq:-43}) independent of the direct specific interaction between colloid and fluid, which can indeed be achieved by using eq. (\ref{eq:-56}). The applicability of the Gibbs adsorption equation is therefore crucial to arrive at separate balance equation for the bulk, as it relies on the existence of an interfacial layer that can simply be 'subtracted'. By eliminating the interfacial term $c\nabla_T\mu_{cs}$ with eq. (\ref{eq:-56}), we obtain

\begin{equation}
c\nabla_T\mu_{c}+\sum_{k\neq 1}n_k^B\nabla_T\mu_{k} =0,\label{eq:-58}
\end{equation}

where $n_k^B=n_{k}-cN_{k}^{\phi}$ is the number of bulk fluid particles per volume. As every colloid occupies a volume $V_c$ of the volume element, $n_k^B$ is related to the bulk density $n_k^b$ of the pure fluid via $n_k^B=n_k^b\left( 1-\varphi\right)$, where $\varphi=cV_{c}$ is the colloidal volume fraction. 

Eq. (\ref{eq:-58}) is independent of the direct interfacial interaction between colloid and fluid and therefore justifies the formulation of a separate microscopic approach that only considers the mutual interaction between colloids in a heat bath. A most general starting point for such a microscopic description is the generalised Fokker-Planck equation \cite{Rub1998} 

\begin{eqnarray}
& & \frac{\partial \mathcal{P}_N}{\partial t}+\sum_i\mathbf{v}_i\nabla_i\mathcal{P}_N+\sum_{ij}\frac{\mathbf{F}_{ij}}{m}\frac{\partial \mathcal{P}_N}{\partial \mathbf{v}_i}\label{eq:-39}\\
= & & \sum_{ij}\frac{\partial}{\partial\mathbf{v}_j}\left[\beta_{ij}\left( \mathbf{v}_j\mathcal{P}_N+\frac{k_{B}T_j}{m}\frac{\partial \mathcal{P}_N}{\partial\mathbf{v}_j}\right)+\gamma_{ij}\mathcal P_N\frac{\nabla_j T}{T_j}\right],\nonumber
\end{eqnarray}

where $\mathcal P_N$ is the N-particle probability distribution of the colloids. The indices $i$ and $j$ run over all colloids inside the volume element, so that $\mathbf F_{ij}$ represents the force that colloid $j$ exerts on colloid $i$. The coefficients $\beta_{ij}$ and $\gamma_{ij}$ are microscopic Onsager coefficients for momentum and heat transfer between colloid $i$ and $j$. Under the assumption that $\gamma_{ij}=0$, the N-particle Smoluchowski equation can be recovered from eq. (\ref{eq:-39}) \cite{Murphy1972}, yielding the result
$\mathbf J_c = -\nabla \Pi/\xi$, where $\Pi$ is the osmotic pressure of the colloids \cite{Dhont2004a,Dhont2004}. The friction coefficient is given by $\xi=6\pi b\eta R/K(\varphi)$, where the mobility factor $K(\varphi)$ accounts for hydrodynamic interactions at finite volume fraction \cite{Batchelor1976}. As this result is obtained with the neglect of $\gamma_{ij}$, we propose the more general form

\begin{equation}
\mathbf J_c = \frac{cT}{\xi}\gamma\left(\varphi \right)\nabla\frac{1}{T} -\frac{1}{\xi}\nabla \Pi,\label{eq:-59}
\end{equation}

where the collective heat coefficient $\gamma\left(\varphi \right)$ disappears when the volume fraction tends to zero. Eq. (\ref{eq:-59}) can be rearranged into the same form as eq. (\ref{eq:-23}) by noticing that $-\nabla\Pi=\nabla P_s^b$. By applying the Gibbs-Duhem equation to a bulk fluid element, the bulk fluid pressure gradient $\nabla P_s^b$ can be expressed as

\begin{eqnarray}
\nabla P_s^b & = & s_s'^b\nabla T+\sum_{k\neq1}n_{k}^b \nabla_T\mu_{k}\\
& = & h_s'^b\frac{\nabla T}{T}+\sum_{k\neq1}n_{k}^b\left\{ \nabla_T\mu_{k} -\mathbf{F}_{k}\right\},\label{eq:-32} 
\end{eqnarray}

where we have used the condition of charge neutrality $\sum_{k\neq 1} n_k^b\mathbf{F}_k=0$ to arrive at eq. (\ref{eq:-32}). Combining eqs. (\ref{eq:-32}), (\ref{eq:-59}) and (\ref{eq:-58}), the flux $\mathbf{J}_c$ can now be written in the Onsager form

\begin{eqnarray}
\mathbf{J}_{c} & = & L_{11}\left(\gamma\left(\varphi \right)-\frac{h_s'^b}{c} \right)\nabla\frac{1}{T} -\frac{L_{11}}{T}\nabla_T\mu_c\\
& & -\frac{L_{11}}{T}\sum_{k\neq1}V_cn_{k}^b\left\{-\nabla_T\mu_{k} +\mathbf{F}_{k}\right\}.\label{eq:-61} 
\end{eqnarray}

By comparing eq. (\ref{eq:-61}) to eq. (\ref{eq:-23}), the 'collective' Onsager coefficients for $\mathbf{J}_c$ can hence be identified as

\begin{eqnarray}
L_{1q}^c & = & \left(\gamma\left(\varphi \right)-\frac{h_s'^b}{c}\right) L_{11},\label{eq:-42}\\
L_{1k}^c & = & -V_cn_k^b L_{11}, \label{eq:-36}\\
\end{eqnarray}

The bulk diffusion flux $\mathbf{J}_c$ can now be combined with the interfacial contribution $\mathbf{J}_{cs}$ to obtain the total colloidal flux $\mathbf{J}$.

\section{The Thermophoretic Flux}

Collecting all derived relations for the Onsager coefficients, given by eqs. (\ref{eq:-27}), (\ref{eq:-74}), (\ref{eq:-42}) and (\ref{eq:-36}), the colloidal flux finally takes the form

\begin{eqnarray}
\mathbf{J} & = & \mathbf{J}_{cs}+\mathbf{J}_c\\
& = & L_{iq}\nabla\frac{1}{T}+\frac{1}{T}\sum_{k}L_{ik}\left\{ -\nabla_T\mu_{k}+\mathbf{F}_{k}\right\},\label{eq:-51}
\end{eqnarray}

where:

\begin{eqnarray}
L_{11} & = & \frac{cT}{\xi},\\
L_{1q} & = & L_{1q}^{cs}+L_{1q}^c=\left(Q_{cs}^*+\gamma\left(\varphi \right)-\frac{h_s'^b}{c}\right) L_{11},\\
L_{1k\neq 1} & = & L_{1k}^{cs}+L_{1k}^{c}=\left( N_k^*+N_k^{\phi}-V_cn_k^b\right) L_{11}. 
\end{eqnarray}

As the solvent ($k=0$) is incompressible, there is no interfacial excess of solvent ($N_0^*=0$) and $\left( dn_0\right) _T=0$. Although the diffusion flux $\mathbf{J}_{c}$ must be balanced by a back-flow of bulk fluid, eq. (\ref{eq:-58}) shows that the corresponding force on a particle of fluid component $k$ is expected to be about $c/n_k$ times smaller than the thermodynamic  force $-\nabla_T\mu_c$. In dilute suspensions ($c\ll n_{k\neq 1}$), it is then reasonable to assume that this back-flow leaves the steady state of the bulk fluid unperturbed. Based on eq. (\ref{eq:-2}), the steady-state distribution of the remaining solutes ($k\neq 0,1$) in the bulk fluid is thus described by $\nabla n_k^b=-n_k^bS_T^k\nabla T$, where $S_T^k$ is the Soret coefficient of solute $k$. The gradients $\nabla_T \mu_k$ and thermoelectric forces $\mathbf{F}_{k}$ in eq. (\ref{eq:-51}) can hence be written as

\begin{eqnarray}
\nabla_T \mu_k & = & \sum_{j \neq 0} \left(\frac{\partial\mu_{k}}{\partial n_j^b}\right)_{P,T}\nabla n_j^b\\
& = & -\nabla T\sum_{j \neq 0,1} n_j^bS_T^j\frac{\partial\mu_{k}}{\partial n_j^b}+\nabla c\frac{\partial\mu_{k}}{\partial c}\label{eq:-66}
\end{eqnarray}

and

\begin{eqnarray}
\mathbf{F}_{k} & = & -z_kV_{T}\frac{\nabla T}{T},\label{eq:-19}
\end{eqnarray}

where $z_k$ is the valence of a particle of component $k$. The thermoelectric potential $V_{T}$ is fixed by the steady-state of the solutes \cite{Majee2011}. To simplify the notation, let us introduce the ratios $\mathcal L_{1k}=L_{1k}/L_{11}$ and $\mathcal L_{1q}=L_{1q}/L_{11}$. By substituting eqs. (\ref{eq:-66}) and (\ref{eq:-19}) into eq. (\ref{eq:-51}), the colloidal flux finally takes the form

\begin{eqnarray}
\mathbf{J} & = & -D\nabla c-cD_T\nabla T,
\end{eqnarray}

where the thermal diffusion coefficient $D_T$ can be identified as

\begin{eqnarray}
\xi D_{T} & = & \frac{\mathcal L_{1q}}{T} -\sum_{k}\mathcal L_{1k}\left\{\sum_{j \neq 0,1} n_j^bS_T^j\frac{\partial\mu_{k}}{\partial n_j^b}-\frac{z_kV_{T}}{T}\right\}\label{eq:-34}
\end{eqnarray}

and the Fickian diffusion coefficient $D$ is given by

\begin{eqnarray}
\xi D & = & c\sum_k \mathcal L_{1k}\frac{\partial\mu_{k}}{\partial c}\label{eq:-40}\\
 & = & \frac{\partial\Pi}{\partial c}+c\sum_{k\neq 1} N_{k}^*\frac{\partial\mu_{k}}{\partial c}.
\end{eqnarray}

From eqs. (\ref{eq:-34}) and (\ref{eq:-40}), it can be seen that the Soret coefficient of the colloids $S_T=D_T/D$ is independent of the friction coefficient. For a separate interpretation of interfacial and collective thermophoresis, it is useful to split $D_T$ up into $D_T=D_T^{cs}+D_T^{c}$, where each term represents the thermal diffusion coefficient of the corresponding flux contribution. From the expressions of $\mathbf{J}_{cs}$ and $\mathbf{J}_{c}$, these coefficients can readily be identified as

\begin{eqnarray}
\xi D_{T}^{cs} & = & \frac{Q_{cs}^*}{T} -\sum_{k\neq 0,1}N_k^*\left\{\sum_{j \neq 0,1} n_j^bS_T^j\frac{\partial\mu_{k}}{\partial n_j^b}-\frac{z_kV_{T}}{T}\right\},\label{eq:-41}\\
\xi D_{T}^{c} & = &\frac{\gamma\left(\varphi \right)}{T} +\frac{1}{c}\frac{\partial \Pi}{\partial T}.\label{eq:-53}
\end{eqnarray}

If the colloids are ideal and point-like ($V_c=0$), we have $\gamma(0)=0$, $\mathcal L_{1k}=\delta_{1k}$ and $n_k^b\partial\mu_{c}/\partial n_k^b=\delta_{1k}k_BT$. The Einstein relation $\xi D=k_BT$ is then recovered from eq. (\ref{eq:-40}). The ideal osmotic pressure is just given by $\Pi=ck_BT$, yielding an ideal thermal diffusion coefficient $\xi D_T=k_B$. In general, both $D_T^{cs}$ and $D_T^{c}$ can depend on the Soret coefficient $S_T^j$ of the solute, meaning that the signs of $D_T^{cs}$ and $D_T^{c}$ do not only depend on whether the specific interactions are attractive or repulsive. In dilute suspensions, colloidal motion is mainly driven by interfacial thermophoresis and the single-particle limit is therefore of particular interest. For a single colloid, the Einstein relation $\xi D=k_BT$ holds and the thermal diffusion coefficient is given by $D_T=D_T^{cs}+k_B/\xi$. The ideal contribution $k_B/\xi$ is usually multiple orders of magnitude weaker than $D_T^{cs}$, so that it can safely be neglected. 

Within the single-particle limit, let us now consider the special case where the fluid only consists of solvent. The Soret coefficient of a colloid is then simply given by $S_T=Q_{cs}^*/(k_BT^2)$. In fact, this result is also commonly used to describe ionic thermophoresis due to hydration \cite{Agar1989}, by treating the ionic solute as a dilute gas of non-interacting, charged particles surrounded by hydration shells. The steady-state of the ionic solute in the bulk is thus governed by
\begin{equation}
\nabla n_j^b+n_j^bS_T^j\nabla T=0\label{eq:-38}
\end{equation}

with an ionic Soret coefficient

\begin{equation}
S_T^j=\frac{Q_{j0}^*+z_jV_T}{k_BT^2}.\label{eq:-26}
\end{equation}

The interfacial heat of transport of the ion $Q_{j0}^*=\mathcal I(h_H)$ is due to the hydration enthalpy density $h_H$ of the surrounding water molecules and the term $z_jV_T$ accounts for the thermoelectric force that directly acts on the ion. It should however be noted that small ions do not necessarily satisfy assumptions 1 and 5 of the hydrodynamic approach, so that deviations of $\mathcal I(h_H)$ from eq. (\ref{eq:-54}) should be expected.
An explicit expression for $V_T$ can further be obtained by multiplying eq. (\ref{eq:-38}) by $z_j$ and summing over all ionic solutes ($j\neq 0,1$), giving

\begin{equation}
V_T=-\frac{\sum_j n_j^b z_j Q_{j0}^*}{\sum_j n_j^b z_j^2},\label{eq:-28}
\end{equation}

where we have also used the condition of charge neutrality $\sum_j z_jn_j^b=0$. Substituting eq. (\ref{eq:-26}) into eq. (\ref{eq:-41}) and noticing again that $n_j^b\partial \mu_k/\partial n_j^b=\delta_{kj}k_BT$ for the ionic gas ($k\neq 0,1$), the thermal diffusion coefficient $D_T^{cs}$ of a single colloid simplifies to 

\begin{equation}
\xi T D_{T}^{cs}=Q_{cs}^*-\sum_{k\neq 0,1}N_k^*Q_{k0}^*.\label{eq:-33}
\end{equation}

For dilute suspensions of charged colloids, eq. (\ref{eq:-33}) shows that the thermal diffusion coefficient $D_T^{cs}$ is directly related to the interfacial heat of transport of colloid and ions, meaning that $D_T^{cs}$ can be evaluated without explicitly determining the thermoelectric potential $V_T$. 

The results that we have derived here make a clear and well-founded statement on the evaluation of transport coefficients in colloidal suspensions, a topic that has been under debate
in recent literature. We will therefore compare our results to other existing theoretical models in the following discussion.

\section{Discussion}

\subsection{Comparison: W\"urger's Force Density for Charged Colloids}

W\"urger et al. \cite{Wurger2010,Morthomas2008} have derived an expression for
the interfacial force density at the surface of a charged colloid in an aqueous
electrolyte solution. The colloidal surface is screened by the ions, leading to the formation of an electric double layer \cite{Debye1923} (Fig. \ref{fig:}). The ions are treated as a non-interacting gas and the local pressure gradient is directly evaluated from the excess pressure $P_{\phi}=P-P_b$ as

\begin{equation}
\nabla P_{\phi}=\nabla\sum_{k\neq 0,1}n_{k}^{\phi}(\mathbf{r})k_{B}T\label{eq:-76}
\end{equation}

with $n_{k}^{\phi}(\mathbf{r})=n_{k}^{b}\left[ \exp\left(-\frac{\phi_{k}(\mathbf{r})}{k_{B}T}\right)-1\right] $. In
our notation, the body force density given by W\"urger reads

\begin{equation}
\mathbf{f}=-\sum_{k\neq 0,1}n_{k}(\mathbf{r})\left(\nabla\phi_{k}(\mathbf{r})-\mathbf{F}_{k}\right)-\frac{1}{2}\epsilon_T\epsilon E^{2}(\mathbf{r})\frac{\nabla T}{T},\label{eq:-30}
\end{equation}

where $\mathbf{F}_{k}$ is the thermoelectric force and $\epsilon_T = \partial\ln\epsilon/\partial\ln T$. The last term in eq. (\ref{eq:-30}) corresponds to the hydration enthalpy density of the polarised solvent (\textit{e.g.} water) in the local electric field $\mathbf{E}$ of the colloid \cite{Landau1960} and should therefore be interpreted as a contribution the the pressure gradient rather than the body force density. With eqs. (\ref{eq:-30}) and (\ref{eq:-76}),
W\"urger's interfacial force density is thus given by
 
\begin{eqnarray}
\vec{\mathcal F}_{\phi} & = & -\sum_{k\neq 0,1}\left(n_{k}(r)\phi_{k}(r)+n_{k}^{\phi}(r)k_{B}T\right)\frac{\nabla T}{T}\label{eq:-25}\\
 &  & -\frac{1}{2}\epsilon_T\epsilon E^{2}(r)\frac{\nabla T}{T}\nonumber \\
 &  & -\sum_{k\neq 0,1}n_{k}^{\phi}(r)\left(k_{B}T\nabla \ln\,n_{k}^{b}-\mathbf{F}_{k}\right).\nonumber
\end{eqnarray}

As expected, we simply have $\nabla_T\mu_k=k_{B}T\nabla \ln\,n_{k}^{b}$ for a non-interacting ionic gas.
The corresponding enthalpy densities at the surface and in the bulk are:

\begin{eqnarray}
h(r) & = &P(r)+\sum_{k\neq 0,1}n_{k}(r)\left(\phi_k(r)+\frac{3}{2}k_BT\right),\\
h_b & = &P_b+\frac{3}{2}k_BT\sum_{k\neq 0,1}n_{k}^b.
\end{eqnarray}

As the partial molar enthalpy of an ideal-gas ion is just
$\bar H_k=\frac{3}{2}k_BT$, the interfacial enthalpy density $h_{\phi}(r)$ of the ions is given by

\begin{eqnarray}
h_{\phi}(r) & = & h(r)-h_b-\sum_{k\neq 0,1}n_{k}^{\phi}(r)\bar{H}_k\\
& = & \sum_{k\neq 0,1}\left(n_{k}(r)\phi_{k}(r)+n_{k}^{\phi}(r)k_{B}T\right).\label{eq:-29}
\end{eqnarray}

With eq. (\ref{eq:-29}), eq. (\ref{eq:-25}) can hence be written in the same form as eq. (\ref{eq:-14}), proving that W\"urger's interfacial force density is in agreement with our more general result. 

\subsection{Comparison: Minimal Models}

Other authors \cite{Dhont2007,Fayolle2005,Bringuier2003,Wurger2006,Dhont2004a} have used different minimal models to derive a force (called internal
or chemical force) from a gradient in a certain potential
$U_{T}$ associated with the colloid.
Most authors have hinted at an interpretation of $U_{T}$ as an excess
chemical potential. The interfacial contribution to $U_{T}$ is usually determined using a 'capacitor' model \cite{Dhont2007}, which considers a Gibbs adsorption process at uniform temperature and pressure:

\begin{equation}
U_{T}^{cs}=-\sum_{k \neq 1}\int N_{k}^{\phi} \left(d \mu_{k}\right)_T=\mu_{cs},\label{eq:-95}
\end{equation}

showing that $U_{T}^{cs}$ indeed corresponds to the interfacial chemical potential (or surface energy) $\mu_{cs}$. Within these minimal models, the colloidal flux is then given
by one of the following forms:

\begin{equation}
\mathbf{J}=-\frac{c}{\xi}\nabla\mu_{exc}-\frac{1}{\xi}\nabla\left(ck_{B}T\right)\label{eq:-15}
\end{equation}

or

\begin{equation}
\mathbf{J}=-\frac{c}{\xi}\nabla\mu_{cs}-\frac{1}{\xi}\nabla\Pi,\label{eq:-6}
\end{equation}

where we recall that $\mu_{exc}=\mu_{cs}+\mu_{cc}$. 

First of all, we notice that none of the above forms accounts for a thermoelectric force $\mathbf{F}_1$. Eqs. (\ref{eq:-15}) and (\ref{eq:-6}) are only equal
if $c\nabla\mu_c=\nabla\Pi$, which is however not a valid thermodynamic identity. Eq. (\ref{eq:-15}) uses a gradient in chemical potential to account for specific interactions but accounts for the ideal contribution with an osmotic pressure gradient, meaning that it neither agrees with our result for $\mathbf{J}_{cs}$, nor with our expression for $\mathbf{J}_{c}$. Eq. (\ref{eq:-6}) contains the appropriate form for $\mathbf{J}_{c}$ with the neglect of $\gamma\left(\varphi \right)$. From comparison to eq. (\ref{eq:-24}), it becomes clear that both forms evaluate $\mathbf{J}_{cs}$ in the H\"uckel limit, which should however not apply to colloidal thermophoresis. 

The general problem with minimal models is that they are purely based on the minimisation of a thermodynamic potential. The form of this potential then automatically imposes certain relations for the Onsager coefficients that should actually be determined based on hydrodynamic and reciprocal arguments, as shown in the previous sections. It is therefore clear that such minimal models cannot properly account for the hydrodynamic character of colloidal thermophoresis.

\section{Conclusion}

We have introduced a well-founded framework for thermophoresis based on the length and time scale separation in colloidal suspensions. This framework justifies the separate evaluation of the interfacial and bulk contribution to the colloidal flux and yields system-specific relations for the Onsager transport coefficients. We have derived a most general expression for the interfacial force density and have shown that thermophoresis cannot be explained by a purely thermodynamic treatment, which only holds in the H\"uckel limit when the colloid is reduced to a point-like particle. The hydrodynamic nature of interfacial thermophoresis is related to irreversible fluid flows in thin boundary layers and is characterised by a non-zero coefficient $L_{1k}^{cs}$. The obtained expression for the thermal diffusion coefficient shows that the strength and direction of thermophoretic motion is not only set by the sign of the specific interaction, but that it also depends on the steady-state of the bulk fluid. We have further shown that the thermal diffusion coefficient of a charged colloid in the presence of an ionic gas can directly be expressed in terms of heat of transport, without an explicit evaluation of the thermoelectric field. Existing limiting cases have also been recovered from our results, showing that our introduced framework draws a clear connection between hydrodynamic and thermodynamic approaches within the theory of NET. 

\section{Acknowledgements}
This work was supported by the Winton Programme for the Physics of Sustainability. D.F. further acknowledges support by the European Union through the European Training Network NANOTRANS Grant 674979, and I.P. acknowledges MINECO and DURSI for financial support under projects FIS2015-67837-P and 2014SGR-922, respectively. We would also like to thank Michael Cates, Ronojoy Adhikari, Alois W\"urger and Jean-Pierre Hansen for stimulating discussions.

\section{Appendix}

\subsection{Thermal Polarization: Computation of the Interfacial Heat of Transport}

Below we derive eq. (\ref{eq:-54}) for the interfacial heat of transport $Q^*_{cs}$ based on Onsager's reciprocity relations by focussing on the heat flux inside the interfacial layer. The computation of $N^*_{k}$ can be treated analogously, by applying the same reciprocal arguments to the flow induced excess transport of component $k$, instead of the heat flux. 

We consider a single colloid subjected to a force $\mathbf F$, moving with a velocity $\mathbf v=\mathbf F/\xi=v{\mathbf {\hat y}}$ through an infinitely large, homogeneous fluid at uniform temperature. ${\mathbf {\hat y}}$ is the unit vector in the direction of $\mathbf F$. The reciprocal relation $L_{1q}^{cs}=L_{q1}^{cs}$ allows us to determine $Q^*_{cs}$ by computing the modified heat flux inside the interfacial layer in the rest frame of the colloid. We restrict ourselves to the case of a stick boundary, although the same procedure may be applied to a slip boundary. For a stick boundary, the fluid flow velocity $\mathbf u_s\left(\mathbf r \right) $ with respect to a spherical colloid moving at $\mathbf v$ can be written as $\mathbf u_s\left(\mathbf r \right)=\mathbf u_{RP}\left(\mathbf r \right)-\mathbf v$, where the contribution $\mathbf u_{RP}\left(\mathbf r \right)$ is described by the Rotne-Prager tensor:

\begin{equation}
\mathbf u_{RP}\left(\mathbf r \right)=\frac{3}{4}\frac{R}{r}\left[
\left(1+{R^2\over 3r^2}\right)\textsf{\textbf{I}}+\left(1-{R^2\over r^2} \right) \mathbf{\hat{r}}\mathbf{\hat{r}}\right]\cdot \mathbf v.
\end{equation}

$\textsf{\textbf{I}}$ is the identity matrix and $\mathbf{\hat{r}}\mathbf{\hat{r}}$ is the dyadic product of the radial unit vector $\hat{r}$. The excess heat transported  by the fluid flow is  given by \cite{Agar1989}

\begin{equation}
Q_{cs}^*\mathbf v=\int_{R}^{\infty}q_{\phi}(r)\mathbf u_s\left(\mathbf r \right)dV.
\end{equation}

Due to the circular symmetry around the line of motion along ${\mathbf {\hat y}}$, only the y-component of $\mathbf u_s$ contributes to the volume integral, so that we can write:

\begin{equation}
Q_{cs}^*v= \int_{R}^{\infty}q_{\phi}(r)\mathbf u_s\left(\mathbf r \right){\mathbf {\hat y}}dV. \label{eq:-47}
\end{equation}

As the interfacial heat density $q_{\phi}(r)$ only depends on the radial distance from the colloidal centre, we can carry out the angular integration of $\mathbf u_s\left(\mathbf r \right){\mathbf {\hat y}}$, yielding $\left\langle \mathbf u_s\left(\mathbf r \right)\cdot {\mathbf {\hat y}}\right\rangle = -v(1-R/r)$. Using this result in eq. (\ref{eq:-47}), we obtain 

\begin{equation}
Q_{cs}^*= -\int_R^\infty 4\pi r^2\left(1-\frac{R}{r}\right) q_{\phi}(r)dr.\label{eq:-21}
\end{equation}

With eq. (\ref{eq:-21}), we have thus recovered the form of $Q_{cs}^*$ for a stick boundary.

Whilst the expression proposed in ref.~\onlinecite{Agar1989} is adequate for particles with no internal degrees of freedom (e.g. atoms), it cannot be used for colloidal particles that can conduct heat internally.  In the latter case, we must account for the fact that the flow-induced heat flux near a particle leads to a thermal polarization that, in its turn, results in an intra-colloidal heat-flux in the direction opposite to the `bare' excess heat flux. The net excess heat flux is the difference between the bare and the intra-colloidal heat fluxes.  Computing the magnitude of the intra-colloidal heat flux is similar to a problem in electrostatics, and is addressed below.

\subsection{Heat Flow in the Boundary Layer Approximation}

To compute the effective, rather than the bare excess heat flux due to flow, we view the colloid in a flow field as 
a spherical heat pump with radius $R$ and thermal conductivity $\kappa_{c}$ embedded in a solvent with thermal conductivity $\kappa_{s}$.
The easiest way to treat the problem of a heat pump in a medium is to consider the total heat flow as a sum of two (fictitious) contributions: the `intrinsic' heat flow $q_0$ and the counterflow $q_1$ induced by the temperature gradient in the sphere. Note that only the total heat flow $Q_0=q_0+q_1$ is observable. Nevertheless, the separation into two fictitious flows is helpful because if the temperature  profile  around the sphere changes, then there will be a real counterflow $\Delta q_1$ and this counterflow is proportional to the {\em change} in the temperature gradient over the sphere.

In the boundary layer approximation ($R\gg\lambda$), the heat flow is generated at the colloidal surface and can be described as resulting from a homogeneous flux density that, by analogy with electrostatics, we denote by ${\mathbf D}_h$:

\begin{equation}\label{eq:Q0D0}
\frac{4}{3}\pi R^3{\mathbf D}_h = {\mathbf J}_h'.
\end{equation}

This heat flux creates temperature gradients inside and outside the colloid. As the temperature must satisfy Poisson's equation, the temperatures inside and outside the colloid are given by

\begin{equation}
T_{in}=A_{in} rP_1(\cos\theta)  +T_0,
\end{equation}
\begin{equation}
T_{out}=A_{out}{P_1(\cos\theta)\over r^2} +T_0,
\end{equation}
where $P_1$ is the first-order Legendre polynomial. Using the continuity condition, we further have
\begin{equation}\label{eq:TinTout2}
A_{out}=R^3 A_{in}.
\end{equation}
If we compute the normal component of the heat flux just outside the sphere, we obtain
\begin{equation}\label{eq:D0plusd0}
D_{out}= 2\kappa_{s} {A_{out}\over R^3} =2\kappa_{s} A_{in},
\end{equation}
where the last equality follows from eq. (\ref{eq:TinTout2}). In addition, the heat flux inside the colloid is given by
\[
D_{in} = - \kappa_{c} A_{in}.
\]
As the total heat flux $D_{out}$ is conserved, we can write
\[
D_{out}=D_h+D_{in}
\]
or
\[
\left(2\kappa_{s}+\kappa_{c}\right) A_{in} = D_h.
\]
It hence directly follows that
\[
D_{out} = {2\kappa_{s}\over 2\kappa_{s}+\kappa_{c}}D_h.
\]
In terms of the integrated heat fluxes, this can alternatively be written as

\begin{equation}
\mathbf J_q'={2\kappa_{s}\over2\kappa_{s}+\kappa_{c}}\mathbf J_h'.
\end{equation}
Hence, the following limiting cases can occur in the boundary layer approximation:
\begin{enumerate}
	\item $\mathbf J_q'=\mathbf J_h'$ if $\kappa_{in}=0$
	\item $\mathbf J_q'=0$ if $\kappa_{in}=\infty$
	\item $\mathbf J_q'=\frac{2}{3}\mathbf J_h'$ if $\kappa_{in}=\kappa_{out}$
\end{enumerate}

In the H\"{u}ckel limit ($R\ll\lambda$), the heat flow through the colloid can be neglected, so that $\mathbf J_q'=\mathbf J_h'$.

\bibliography{Theory_references}

\end{document}